\documentclass[twocolumn,showpacs,preprintnumbers,amsmath,amssymb]{revtex4}
\usepackage[dvips]{graphicx}
\usepackage{setspace}
\usepackage{amsmath}
\usepackage{amsfonts}
\usepackage{amssymb,color} 
\usepackage{graphicx}
\usepackage{bm} 
\usepackage{psfrag}
\usepackage[utf8]{inputenc}
\begin{document}

\title{Glassy dynamics and hysteresis in a linear system of orientable hard rods}

\author{Jeferson J. Arenzon}
\email{arenzon@if.ufrgs.br}
\affiliation{Instituto de F\'\i sica and
National Institute of Science and Technology for Complex Systems,
\\ Universidade Federal do Rio
Grande do Sul \\ C.P. 15051, 91501-970, Porto Alegre, RS, Brazil}

\author{Deepak Dhar}
\email{ddhar@theory.tifr.res.in}
\affiliation{Department of Theoretical Physics \\
Tata Institute of Fundamental Research \\
Homi Bhabha Road, Mumbai, 400005, India}

\author{Ronald Dickman}
\email{dickman@fisica.ufmg.br}
\affiliation{Departamento de F\'{\i}sica, Instituto de
Ci\^encias Exatas, and National Institute of Science and Technology for Complex Systems,\\
Universidade Federal de Minas Gerais \\
C.P. 702, 30123-970, Belo Horizonte, MG, Brazil }

\date{\today}

\begin{abstract}
We study the dynamics of a one-dimensional fluid of orientable hard rectangles
with a non-coarse-grained microscopic mechanism of facilitation.
The length occupied by a rectangle
depends on its orientation, which is coupled to an external field.
The equilibrium properties of our model are
essentially those of the Tonks gas, but at high densities, the
orientational degrees of freedom become effectively frozen due to jamming.  
This is a simple analytically tractable model of glassy phase. Under a cyclic variation
of the pressure, hysteresis is observed.
Following a pressure quench, the orientational persistence exhibits a
two-stage decay characteristic of glassy systems.
\end{abstract}
\maketitle

\section{Introduction}
Almost all liquids, if cooled sufficiently fast, form a
glassy structure; much effort has been directed towards understanding
this phenomenon.  Monte Carlo simulations have shown that systems with
purely hard-core interactions can describe qualitatively much of the
observed phenomenology of the glass transition, e.g., the very fast rise
of relaxation times, and the absence of an associated latent heat.
Theoretical analysis of such models is hampered by an incomplete
understanding of the equilibrium (fluid-solid) phase transition. The
best studied hard-core model is a system of hard spheres, which has
experimental realizations in colloidal, granular and other systems~\cite{PuMe86,ArTs06,PrSeWe07}.
In many cases size dispersion, or other built-in complexity, is
introduced in order to avoid crystallization and thereby observe a glass
transition~\cite{SaKr00,PaZa05,PaZa06,Zamponi07b}. For monodisperse systems, there is no transition
in one dimension~\cite{Tonks36}, while two and three dimensional systems are highly
prone to crystallize. 
In higher dimensions~\cite{SkDoStTo06,MeFrCh09,ZaVaSaPoCaPu09,ChIkMeMi10},
nucleation rates are low and the glass state is more easily attained.
Indeed, in the limit of very high spatial
dimensionality, there seems to be an ideal glass transition~\cite{PaZa06,ScSc10},
although how far this extends down to lower dimensions is still debated~\cite{Tarzia07}.

Hard core potentials need not  be spherically symmetric. On a
lattice, for example, where the symmetry is discrete, the behavior
strongly depends on the dimension of the system, the lattice
structure, and the exclusion range (see
\cite{Panagiotopoulos05,FeArLe07} and refs. therein).

Here we study the dynamic properties of a one-dimensional system of classical 
hard rectangles, with only two orientations allowed, horizontal and vertical. These will be called ``rods" in what follows.
Classical linear fluids have been extensively
studied over the last
decades~\cite{Tonks36,Gursey50,Baxter65,LiMa66,CaRu69,Bloomfield70,Percus76,Percus82,Marko89,Davis90,DuHu03,FeLeAr07,TrPa08};
the case of elongated rods with orientational
freedom has also been analyzed~\cite{CaRu69,FrRu73,FuLo79,BeKr06,KaKa09a,KaKa09b,GuVa10}.
While the equilibrium properties of this class of models are well
understood, their dynamic properties, in particular those related
to the glass transition, have received somewhat limited
attention~\cite{Bowles00,StLuMe02,StDe02,BoSa06,DhLe10}.
The simple, one-dimensional model considered here, reproduces (at least in part)
the glassy phenomenology with an explicit (non-coarse-grained) microscopic
mechanism of facilitation.
In analogy with closely related models in
which the constraints are kinetic instead of
geometric~\cite{FrAn84,KoAn93,RiSo03}, due to a spatial and temporal
coarse-graining, here there is no thermodynamic transition and the
only nontrivial behavior is kinetic. In the present context this represents an
advantage, in that the relaxation is not complicated by
critical slowing down or metastability.

At short time scales, the glassy phase can be modelled as a metastable phase, in restricted thermal equilibrium. 
An exact calculation of the properties 
of such metastable  states,  including the equation of state,  
has been made recently for a  toy model~\cite{Dhar02,DhLe10}.
 In these papers, the ergodicity is explicitly broken, and one assumes that in  the glassy phase 
some transition rates are exactly zero.  The simple model considered here
provides an extension of the treatment in Ref.~\cite{DhLe10}
to include  the description of slow evolution of the macroscopic  glassy state at longer time scales.
In our model, ergodicity is not explicitly broken, and  the dynamics is capable of bringing the system to
equilibrium, but at high pressures, the relaxation of orientations
becomes so slow that these variables are effectively {\it frozen} over any reasonable time scale.
Under steady increase of pressure, they never fully relax.
Macroscopic properties in the frozen regime are found to depend on history, in particular,
on the rate at which the pressure is increased.

The structure of the paper is as follows.  The next section
introduces the model while Sec.
\ref{sec.equilibrium} reviews its equilibrium properties.
 We describe the dynamic behavior in section
\ref{sec.dynamics}. Sec. \ref{sec.conclusions} contains some 
concluding remarks.

\section{Model}
\label{sec.model}

\begin{figure}[ht]
\includegraphics[width=6cm]{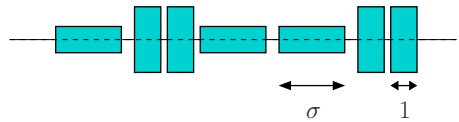}
\caption{Example of a rod configuration.}
\label{fig.rods3}
\end{figure}

We consider a system of $N$ rigid rods of
length $\sigma \geq 1$ and unit width on a line of length $L$.
Each rod is described by the
variables $(x_i,S_i)$, where $x_i\in [0,L)$ is the position of the center of mass
and $S_i$ denotes its orientation ($0$ for horizontal, 
1 for vertical). Rods in state $S_i=0$ occupy a length $\sigma$ and
those in state $S_i=1$ occupy a unit length (see Fig.~\ref{fig.rods3}).
Since the order of the particles cannot change, we
take $x_1 < x_2 < \ldots < x_N$.
For convenience, we define
$x_0 \equiv 0$ and $x_{N+1}\equiv L$; there are no orientational
degrees of freedom associated with these variables.
We define inter-rod distances, $y_i = x_{i+1} - x_i$, so that $\sum_{i=0}^N y_i = L$.
When all rods are constrained to have the same orientation or,
equivalently, $\sigma=1$, the original  Tonks gas~\cite{Tonks36} is
recovered. Without this restriction, the model is
analogous~\cite{Onsager49} to a binary mixture of rods (with
conservation of the total particle number, but not the number of
each species separately). The particles are
subject to an external field $h'$, coupled to the
orientations $\{S_i\}$, so that the potential energy takes the form
\begin{equation}
{\cal H} = 
\sum_i \phi_{\scriptstyle\rm hc}(y_i) + h'\sum_i S_i,
\end{equation}
where the first term denotes hard core interactions and the sums
extend over the $N$ rods; for $h'>0$ the horizontal orientation is favored.

The hard-core interaction
between the $i$th rod and its right neighbor is
$$
\phi_{\scriptstyle\rm hc}(y_i) =
\begin{cases}
0, & y_i\geq a_i \\
\infty, &  y_i< a_i,
\end{cases}
$$
where $a_i$ is the minimum distance between the centers of rods
$i$ and $i+1$, given by
\begin{equation}
a_i = \sigma + \frac{1}{2}(S_i+S_{i+1})(1-\sigma).
\label{mindist}
\end{equation}
An important quantity is the mean free volume per particle,
$v_f= v-\sigma+(\sigma-1)m$, where $v=\rho^{-1}= L/N$ and $m$ is the
fraction of rods in the vertical orientation, or ``magnetization''.

\begin{figure}[ht]
\includegraphics[width=6cm]{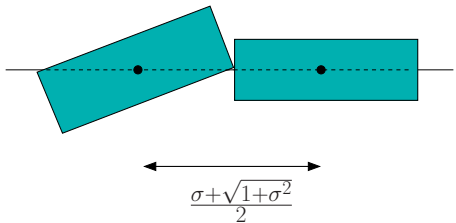}
\caption{Geometrical constraint associated with an orientational transition. Analogous
restrictions apply also for the other possible orientations.}
\label{flip}
\end{figure}

We define a local, continuous-time,
stochastic dynamics for the positions (or equivalently, the separations
$y_i$), and orientations. The time evolution is  Markovian, and satisfies the detailed balance
condition.
Each rod executes unbiased diffusive motion. The wall
at $x_{N+1} = L$ also undergoes diffusive motion,  but is subject to a bias,
with ratio of jump rate that increase $L$ by $\delta L$ to the backward transition is 
$\exp( -  p \delta L)$. 
The rods can also change their orientation, rotating about its center of mass. 
While the orientation varies continuously 
during a transition, we suppose the latter occurs so rapidly that orientations other than 
vertical or horizontal may be neglected. 
In addition, the dynamics includes an important constraint on
orientational transitions: particle $i$ can only change its
orientation if its neighbors to the left and right are sufficiently
far away that its rotation (through 90$^{\rm o}$) is not blocked geometrically
(see Fig.~\ref{flip}).
In order for particle $i$ to change its orientation, it is necessary that
\begin{equation}
y_i>\frac{\sigma+S_{i+1}(1-\sigma)+r}{2}
\label{eq.condition}
\end{equation}
where $r\equiv \sqrt{1+\sigma^2}$ is the rod diagonal.
The analogous relation for $y_{i-1}$ must be satisfied as well. 
Note that these conditions
depend on the states of the neighboring rods
($i\pm 1$), but not on $S_i$ itself. Without the nonoverlapping
constraint, $r$ is replaced by $\sigma$ in the above equation.
Any  violation of the excluded volume condition is rejected.
These transitions are accepted
in accordance with the Metropolis criterion, that is, with probability
$\min [1,\exp(-\beta\Delta {\cal H})]$. 
One Monte Carlo step (MCS) consists
in an attempt to update all degrees of freedom (i.e., all positions and orientations, and
the volume).

Our Monte Carlo simulations were performed
on systems of $N=1000$ rods; and results represent averages over
100 (or more) independent realizations.

\section{Equilibrium Properties}
\label{sec.equilibrium}

\begin{figure}[ht]
\includegraphics[width=8cm]{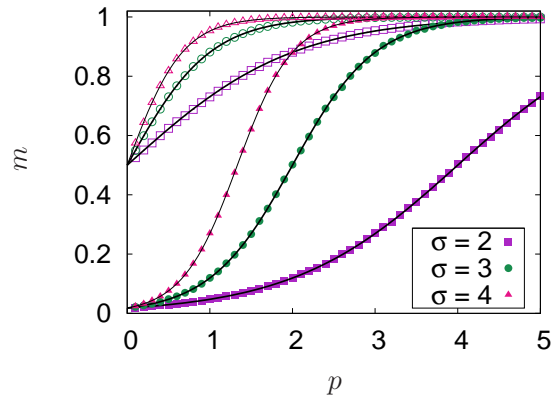}
\caption{Equilibrium fraction of vertical rods as a function of pressure
$p$, for $h=0$ and 4, (empty and filled symbols, respectively).
Points are simulation results,
while the lines correspond to Eq.~(\ref{fracvert}). When $p$ is
large enough, all rods are aligned vertically.
}
\label{fig.mp}
\end{figure}

The equilibrium properties of the model
are indeed very simple.
The canonical configurational partition function is given by
\begin{equation}
Z_N(h,L) = 
\sum_{\{S_i\}} e^{-h \sum_i S_i}
\int_R dy_1 \cdots dy_N \label{Z}
\end{equation}
where
$h=\beta h'$ and the subscript $R$
denotes the restrictions $\sum_{i=1}^N y_i = L$
and $y_i \geq a_i$, where the minimum distances $a_i$ are defined in
Eq.~(\ref{mindist}).  The usual factor of $1/N!$ is absent due to the fixed order
of the particles on the line. Introducing the variables $z_i = y_i -
a_i$, (with $z_0 \equiv x_1$ and $z_N \equiv L - x_N$), we have,
\begin{eqnarray*}
Z_N(h,L) &=&
\sum_{\{S_i\}} e^{-h \sum_i S_i}
\int_0 dz_0 \cdots \int_0 dz_N \;\nonumber \\ &&
\times \delta \left[ \sum_{i=0}^N z_i - L +
N\sigma - (\sigma-1)\sum_{i=1}^N S_i \right].
 \label{Z1}
\end{eqnarray*}
We study the system in the constant-pressure ensemble;
the partition function is
\[
Y_N(h,p) = \int dL e^{-p L} Z_N(h,L) ,
\]
where $p$ denotes the pressure divided by $k_B T$.  A simple calculation yields
\begin{equation}
Y_N (h,p) =  
e^{-p N \sigma} p^{-N}(1+ \kappa)^N
\end{equation}
with $\kappa \equiv \exp [(\sigma-1)p - h]$.
The Gibbs free energy per particle is given by $g = -(N \beta)^{-1}
\ln Y$, where, in the thermodynamic limit,
\begin{equation}
\lim_{N \to \infty} \frac{1}{N} \ln Y_N = -\sigma p + \ln (1+\kappa)
-\ln p . 
\label{tlim}
\end{equation}
In this limit the volume per particle is,
\begin{equation}
v(p,h) = -\frac{1}{N} \frac{\partial \ln Y}{\partial p}
= \sigma - \frac{(\sigma -1) \kappa}{1+\kappa} + \frac{1}{p},
\label{volperpart}
\end{equation}
while the fraction of vertical rods is,
\begin{equation}
m_{eq}(p,h) = -\frac{1}{N} \frac{\partial \ln Y}{\partial h} =
\frac{\kappa}{1+\kappa}. \label{fracvert}
\end{equation}
Eqs. (\ref{volperpart}) and (\ref{fracvert}) imply the relation
\begin{equation}
v = \frac{1}{p} + m_{eq} + \sigma (1-m_{eq}),
\label{vmp}
\end{equation}
which implies that the free volume per particle is $v_f = 1/p$, as in the Tonks
gas.  It is worth noting than in equilibrium, the variables $\{y_i \}$ and
$\{S_i \}$ are all mutually independent.
The behavior of $m$ as a function of the pressure for
several values of $\sigma$ and $h$ is illustrated in
Fig.~\ref{fig.mp}.

\section{Time-dependent Properties}
\label{sec.dynamics}

In this section we study the kinetics of the magnetization $m$ and molecular
volume $v$.
We assume that the diffusive relaxation  is much faster than  the 
orientational relaxation.
Then the displacement degrees of freedom may be assumed to be in thermal equilibrium. 
At high pressures, most update attempts for orientation change fail, as they are blocked. 
Suppose first that the orientations are fixed.  
In this case, the different sectors in the pico-canonical  ensemble are specified by orientation of each rod. Different sectors with the same number of vertical rods are macroscopically equivalent. Within a sector, the displacement degrees of freedom are assumed to be in equilibrium.
Then the translational dynamics
will bring the system to a constrained equilibrium distribution, in which the
distances $z_i$ are independent, exponentially distributed random variables with mean
$1/p$, and the mean length of the system is
$\langle L \rangle = N[1 + (1-m) \sigma] + (N+1)/p]$, where $m$ is the fraction of
vertically oriented particles, not necessarily equal to the equilibrium value
$m_{eq}$.

Now, allowing the orientations to fluctuate, an equation of motion for $m(t)$ can be
derived if we assume that the translation dynamics is rapid, so that between any pair of
successive orientational transitions, the interparticle distances attain the constrained equilibrium
distribution mentioned above. Under this hypothesis, the particle orientations are
mutually independent, as in equilibrium (we assume as well that the orientations are
initially uncorrelated).

Consider, for example, a transition from $S_i=0$ to 1. This would be allowed only if the 
two gaps on the two sides of the rod are large enough. 
In the restricted equilibrium ensemble, separations are independent,
exponentially distributed random variables, $P(y)=p\exp(-py)$, and
the probability that a hole larger than $w=(r-\sigma)/2$ appears at one
of its sides is $P(v_f>w)=\exp[-p(r-\sigma)/2]$. Given
that such gaps must exist at both sides and
that this flip is against the field, we see that the effective transition rate
is proportional to $\exp[-p(r-\sigma)-h]$, and that it is independent of
the state of the neighboring rods. Thus we may write the transition rate
for $S_i=0$ to 1 as
\begin{equation}
\gamma_+ = \gamma \exp[- p(r -\sigma) - h],
\label{gammaplus}
\end{equation}
where $\gamma$ is an arbitrary attempt rate, independent of $p$ and $h$.
On the other hand, if the transition is from 1 to 0, $w=(r-1)/2$ and
\begin{equation}
\gamma_- = \gamma \exp[- p(r - 1)].
\label{gammamin}
\end{equation}
Note that only $\gamma_+$ depends on $h$ and that these rates
satisfy detailed balance. 
For small values of the pressure, free space is abundant and the slowest
process is a flip against the field, so that the larger time scale
involved is given by $\gamma_+^{-1}$.
On the other hand, when the pressure is large enough, the production
of large enough holes is the dominant slow process and the
relevant characteristic time  now scales as $\gamma_-^{-1}$.

The evolution of $m(t)$ is governed by
\begin{equation}
\frac{dm}{dt} = \gamma_+ (1-m) - \gamma_- m \equiv - \Gamma m + \gamma_+ ,
\label{dmdt}
\end{equation}
where $\Gamma = \gamma_+ + \gamma_-$. If the rates are time independent,
then letting $\phi = m - m_{eq} = m - \gamma_+/\Gamma$, we have
\begin{equation}
\frac{d\phi}{dt} = -\Gamma \phi,
\end{equation}
showing an exponential approach to equilibrium.

The system is driven out of equilibrium if the pressure (or the external field, $h$)
is time-dependent.
Suppose that $\gamma_+$ and $\gamma_-$ depend on time through the pressure.  Then we have
\begin{equation}
m(t) = e^{-{\cal G}(t)}
\left( m_0 + \int_0^t ds \; \gamma_+ (s) \; e^{{\cal G}(s)} \right),
\label{mt}
\end{equation}
where
\begin{equation}
{\cal G}(t) = \int_0^t dt' \, \Gamma(t').
\end{equation}

A particularly interesting example is that of a system initially in equilibrium at pressure $p_0$, and
subject to a pressure that increases linearly with time, $p(t) = p_0 + \lambda t$ for $t > 0$,
where $\lambda$ is the annealing rate. 
In this case,
\begin{align}
{\cal G}(t) = \frac{\gamma}{\lambda} &
\left[ e^{-h-p_0(r-\sigma)} \frac{1 - e^{-\lambda (r-\sigma)t}}{r-\sigma} \right.\nonumber\\
& \left.  + e^{-p_0(r-1)} \frac{1 - e^{-\lambda (r-1)t}}{r-1} \right]
\end{align}
so that
\begin{equation}
\lim_{t \to \infty} {\cal G}(t) = \frac{\gamma}{\lambda}
\left[\frac{e^{-h-p_0(r-\sigma)}}{r-\sigma}
+ \frac{e^{-p_0(r-1)}}{r-1} \right],
\end{equation}
which is finite for $\lambda > 0$. Inserting this result in Eq.~(\ref{mt}),
we see that the initial magnetization $m_0$
is not ``forgotten" even when $t \to \infty$.
It is easily verified that memory of the initial magnetization persists
for a pressure increase of the form $p(t) = p_0 + \lambda t^{\alpha}$,
for any positive values of $\lambda$ and $\alpha$.

\begin{figure}[ht]
\includegraphics[width=8cm]{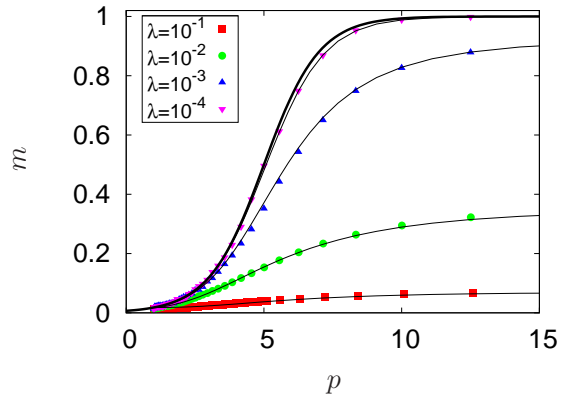}
\caption{Fraction $m$ of vertical rods versus pressure
$p$ for $h=5$, $\sigma=2$, $\gamma=1$ and several values of the rate of pressure increase, $\lambda$.
From top to bottom, $\lambda=0$ (equilibrium, bold line), $10^{-3}$, $10^{-2}$, and $10^{-1}$.
Points: simulation; solid lines: theory, Eq.~(\ref{mt}).}
\label{p_ann_h5_lambda}
\end{figure}

\begin{figure}[ht]
\includegraphics[width=8cm]{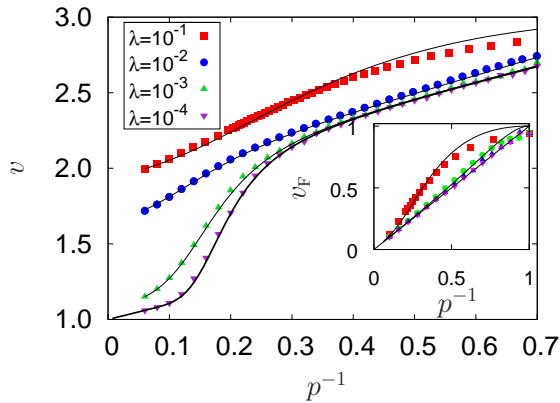}
\caption{Volume per molecule $v$ versus pressure
$p$ for the same parameters as in Fig. \ref{p_ann_h5_lambda} during the pressure
annealing. Points represent
the simulation data while the solid lines are the improved theoretic predictions
based on Eq.~(\ref{dvfdt}) (with $\gamma\simeq 0.3$).
The
lines for $\lambda=10^{-4}$ and $\lambda=0$ (equilibrium, bold line)
 are indistinguishable at this scale. Inset: free volume $v_F$. Analogous
 deviations from equilibrium are again seen for large values of $\lambda$.}
\label{p_ann_h5_lambda2}
\end{figure}

Examples of $m(t)$ (for pressure increasing linearly with time) 
are shown in Fig.~\ref{p_ann_h5_lambda} for $p_0=1$, $h=5$, $\sigma = 2$,
and $\gamma = 1$.  For $\lambda = 10^{-4}$, the difference between $m(p)$ and the equilibrium
result is small.
For larger rates of pressure increase, on the other hand, there are marked differences
between the final value of $m$ and the equilibrium result.  Although $m(t)$ is well
described by Eq.~(\ref{mt}), as can be seen by the excellent agreement with the simulation,
the same does not occur for the molecular volume, $v(p)$, and the free volume,
at larger rates of pressure increase $\lambda$.
The reason is that in this case
the rate of pressure increase is large enough so that we can no longer treat the 
interparticle distances as in instantaneous thermal equilibrium, on the time scale of the
orientational relaxation.  This effect can be incorporated in an approximate manner in the
theory if we assume that the free volume per molecule follows a relaxational dynamics, that is,
\begin{equation}
\frac{d v_F}{dt} = - \Gamma_v \left( v_F - \frac{1}{p} \right)
\label{dvfdt}
\end{equation}
To evaluate the transition rates $\gamma_+$ and $\gamma_-$ we require the probability density $p(z)$.
The simplest hypothesis is that $p(z)$ is exponential, as in equilibrium, but with the mean free
volume $v_F$ in place of its equilibrium value, $1/p$, so that $p(z) = v_F^{-1}\exp(-z/v_F)$.
Then the evolution of the magnetization is given by Eq.~(\ref{dmdt}), with transition rates as
in Eqs.~(\ref{gammaplus}) and (\ref{gammamin}), but with $p$ replaced by $1/v_F$.  With an
appropriate choice of the relaxation rate $\Gamma_v$, this simple theory yields reasonable agreement
with simulation results at larger quench rates, as is shown in
Fig.~\ref{p_ann_h5_lambda2}. Some deviations between the theoretical prediction and simulations
are evident at the highest quench rate, for smaller pressures; this is not surprising given the
simplifications introduced.

\begin{figure}[ht]
\includegraphics[width=8cm]{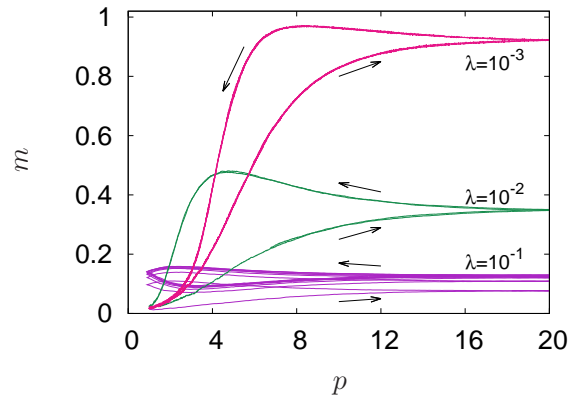}
\caption{Fraction of vertical rods under cyclic variation of the pressure, showing hysteresis
loops. For the highest rate, $\lambda=0.1$, the loop only closes after several cycles.
Parameters: $h=5$, $\sigma=2$; the pressure varies between 1 and 20 at
rates $\lambda$ as indicated.}
\label{histerese}
\end{figure}

In addition to the ``freezing out'' of the orientational degrees of freedom under a steady
pressure quench, the system exhibits interesting hysteresis effects under an oscillatory
pressure.  Hysteresis loops obtained through numerical simulation are illustrated in
Fig.~\ref{histerese}, for various values of $\lambda = |dp/dt|$ in triangle-wave cycles
of pressure variation. Similar results are obtained via numerical integration of
Eq.~(\ref{dmdt}), assuming rapid equilibration of the free volume. Notice that for
larger rates, the system describes a sequence of irreversible loops before
entering a reversible one.  For even larger rates than those shown in the
figure, we find a greater number of irreversible loops, analogous
to those obtained in compaction experiments of rods under vibration~\cite{ViLaMuJa00}.

It is worth noting that an external field $h$ is not required to observe freezing.
Even with $h=0$, the rapid reduction in the transition rate $\gamma_+$, Eq.~(\ref{gammaplus}),
with increasing pressure ensures that the orientational degrees of freedom cannot equilibrate.
Inhibition of orientational relaxation is greatest for $\sqrt{1 + \sigma^2} - \sigma$ as large as
possible, i.e., for $\sigma$ tending to unity.  (Of course this tends to reduce the excess of
$v$ over its equilibrium value.)
Thus, for $h=0$, $\sigma = 1.1$, and other parameters as above, one finds
$m_\infty \equiv \lim_{t \to \infty} m(t) = 0.620$ if the pressure increases at a rate of $\lambda = 1$, and
$m_\infty = 0.5409$ for $\lambda = 10$.  (As $\lambda$ is increased, $m_\infty $ approaches the
initial magnetization, equal to $1/2$ for $h=0$.)

\begin{figure}[ht]
\includegraphics[width=9cm]{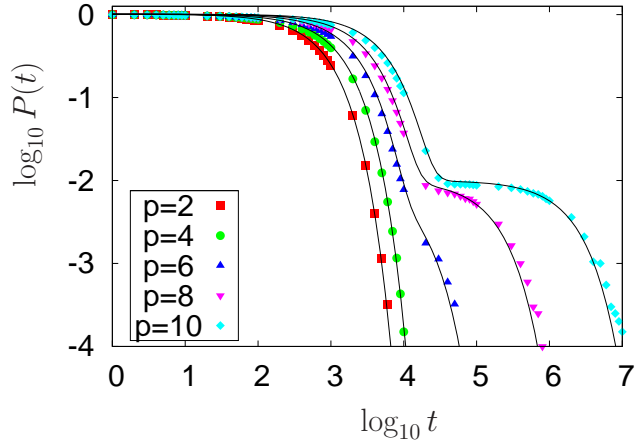}
\caption{Persistence $P(t)$ as a function of time as the system pressure is quenched
from 1 to $p$ at $t=0$, for $\sigma=2$ and $h=5$. When $p$ is small, the decay is exponential
while for larger values, after an initial exponential decay, the
persistence develops a plateau whose height is close to the initial magnetization
$m_0$ and whose width increases with $p$. The
solid lines are from Eq.~(\ref{eq.per}).}
\label{fig.quench_per}
\end{figure}

If instead of a smooth annealing, the system is suddenly
quenched from low to high pressure, a two-step decay typical
of glassy behavior is observed in the orientational persistence 
(the fraction of rods that did not flip since $t=0$). 
Fig.~\ref{fig.quench_per} shows
the results for several values of the final pressure after the system
is quenched from an equilibrium state at $p_0=1$. Whatever the value of
the final pressure, the initial decay of the persistence function is exponential
and ruled by the against the field flip of rods that are horizontal at $t=0$,
so that $P(t)\simeq \exp(-\gamma_+t)$ in this regime. As the final pressure is increased,
the flip from 1 to 0 becomes slower and
a plateau develops, after the initial fast decay, whose width increases
with the final pressure. The height of this plateau is on the order
of the initial magnetization, $m_0$.  (Its precise value is
slightly smaller than $m_0$ since some of the up rods will have already flipped during
the initial fast regime). When
the final pressure is large, it takes
a long time for a rod to flip, since enough space must be freed, which
requires a cooperative rearrangement of the neighboring rods. This happens with a
rate proportional to $\gamma_-$. Thus, the data in Fig.~\ref{fig.quench_per}
are well fit by a sum of exponentials,
corresponding to the fast and slow processes in the model:
\begin{equation}
P(t) \simeq {\rm e}^{-\gamma_+t} + m_0{\rm e}^{-b\gamma_-t}
\label{eq.per}
\end{equation}
where, for the parameters of Fig.~\ref{fig.quench_per}, $b\simeq 0.4$ and
$m_0\simeq 0.01$.  
The coefficient of the second term, 0.01, is the  height of the plateau
and roughly corresponds to the initial magnetization. Thus, starting with
an even smaller initial pressure, and thus a larger magnetization,
the plateau can be tuned to higher values. The width of the plateau
increases with the final value of the pressure and diverges
as $p\to\infty$. This diverging relaxation time, $\gamma_-^{-1}$, is
related to the increasing length of the cooperative region~\cite{KaKa09b}.
  Notice that, at variance with other models
for the glass transition, the slow relaxation is not associated with stretched exponentials.

\section{Conclusions}
\label{sec.conclusions}

One-dimensional systems with short-range interactions, such as the model
studied here, do not exhibit a phase transition at finite temperature
and pressure in equilibrium.
A jamming transition involving certain degrees of freedom
may nevertheless occur~\cite{KaKa09b}, with a
diverging length scale, as the control parameter (inverse pressure or
temperature) goes to zero. Here we study a
simple, geometric model on the line, for which analytical results
for both the statics and the dynamics may be obtained and
compared with numerical simulations, with excellent
agreement.  This system of hard rods is subject to geometric
constraints that prohibit a rod changing its orientation if
the distance from its neighbors is too small (i.e., in the
absence of sufficient free volume).
In our model, the time spent in transit between the two orientations 
is assumed to be small, and the orientation degree was taken to be 
a discrete variable. It is straightforward to extend the discussion 
to the case where we allow  continuous orientations, although no qualitative 
differences are expected.

In the
presence of an external field that disfavors the vertical
position, our model may be seen as an
in-layer description of a higher-dimensional system.  In a dense
system, the vertical position will be disfavored due to
excluded-volume interactions with the rods in the
neighboring layers. Thus the external field may be interpreted
as an effective interaction to take into account the remaining
dimensions.

Despite its simplicity, the model presents several properties
characteristic of the glass transition, such as annealing rate
dependence and two-step relaxation. 
The dynamical behavior can be understood
in terms of the two microscopic reorientation
processes involved.

\begin{acknowledgments}
The work  of JJA and RD is  partially supported by the Brazilian
agencies FAPERGS, FAPEMIG, CAPES and CNPq. RD was partially 
supported by CNPq under project 490843/2007-7.
DD  was partially supported by the Department of
Science and Technology, Government of India under the project
DST/INT/Brazil/RPO-40/2007.
\end{acknowledgments}


\begin{thebibliography}{45}
\expandafter\ifx\csname natexlab\endcsname\relax\def\natexlab#1{#1}\fi
\expandafter\ifx\csname bibnamefont\endcsname\relax
  \def\bibnamefont#1{#1}\fi
\expandafter\ifx\csname bibfnamefont\endcsname\relax
  \def\bibfnamefont#1{#1}\fi
\expandafter\ifx\csname citenamefont\endcsname\relax
  \def\citenamefont#1{#1}\fi
\expandafter\ifx\csname url\endcsname\relax
  \def\url#1{\texttt{#1}}\fi
\expandafter\ifx\csname urlprefix\endcsname\relax\def\urlprefix{URL }\fi
\providecommand{\bibinfo}[2]{#2}
\providecommand{\eprint}[2][]{\url{#2}}

\bibitem[{\citenamefont{Pusey and van Megen}(1986)}]{PuMe86}
\bibinfo{author}{\bibfnamefont{P.~N.} \bibnamefont{Pusey}} \bibnamefont{and}
  \bibinfo{author}{\bibfnamefont{W.}~\bibnamefont{van Megen}},
  \bibinfo{journal}{Nature} \textbf{\bibinfo{volume}{320}},
  \bibinfo{pages}{340} (\bibinfo{year}{1986}).

\bibitem[{\citenamefont{Aranson and Tsimring}(2006)}]{ArTs06}
\bibinfo{author}{\bibfnamefont{I.~S.} \bibnamefont{Aranson}} \bibnamefont{and}
  \bibinfo{author}{\bibfnamefont{L.~S.} \bibnamefont{Tsimring}},
  \bibinfo{journal}{Rev. Mod. Phys.} \textbf{\bibinfo{volume}{78}},
  \bibinfo{pages}{641} (\bibinfo{year}{2006}).

\bibitem[{\citenamefont{Prasad et~al.}(2007)\citenamefont{Prasad, Semwogerere,
  and Weeks}}]{PrSeWe07}
\bibinfo{author}{\bibfnamefont{V.}~\bibnamefont{Prasad}},
  \bibinfo{author}{\bibfnamefont{D.}~\bibnamefont{Semwogerere}},
  \bibnamefont{and} \bibinfo{author}{\bibfnamefont{E.~R.} \bibnamefont{Weeks}},
  \bibinfo{journal}{J. Phys.: Condens. Matt.} \textbf{\bibinfo{volume}{19}},
  \bibinfo{pages}{113102} (\bibinfo{year}{2007}).

\bibitem[{\citenamefont{Santen and Krauth}(2000)}]{SaKr00}
\bibinfo{author}{\bibfnamefont{L.}~\bibnamefont{Santen}} \bibnamefont{and}
  \bibinfo{author}{\bibfnamefont{W.}~\bibnamefont{Krauth}},
  \bibinfo{journal}{Nature} \textbf{\bibinfo{volume}{405}},
  \bibinfo{pages}{550} (\bibinfo{year}{2000}).

\bibitem[{\citenamefont{Parisi and Zamponi}(2005)}]{PaZa05}
\bibinfo{author}{\bibfnamefont{G.}~\bibnamefont{Parisi}} \bibnamefont{and}
  \bibinfo{author}{\bibfnamefont{F.}~\bibnamefont{Zamponi}},
  \bibinfo{journal}{J. Chem. Phys.} \textbf{\bibinfo{volume}{123}},
  \bibinfo{pages}{144501} (\bibinfo{year}{2005}).

\bibitem[{\citenamefont{Parisi and Zamponi}(2006)}]{PaZa06}
\bibinfo{author}{\bibfnamefont{G.}~\bibnamefont{Parisi}} \bibnamefont{and}
  \bibinfo{author}{\bibfnamefont{F.}~\bibnamefont{Zamponi}},
  \bibinfo{journal}{J. Stat. Mech.} \textbf{\bibinfo{volume}{P03017}}
  (\bibinfo{year}{2006}).

\bibitem[{\citenamefont{Zamponi}(2007)}]{Zamponi07b}
\bibinfo{author}{\bibfnamefont{F.}~\bibnamefont{Zamponi}},
  \bibinfo{journal}{Phil. Mag.} \textbf{\bibinfo{volume}{87}},
  \bibinfo{pages}{485} (\bibinfo{year}{2007}).

\bibitem[{\citenamefont{Tonks}(1936)}]{Tonks36}
\bibinfo{author}{\bibfnamefont{L.}~\bibnamefont{Tonks}},
  \bibinfo{journal}{Phys. Rev.} \textbf{\bibinfo{volume}{50}},
  \bibinfo{pages}{955} (\bibinfo{year}{1936}).

\bibitem[{\citenamefont{Skoge et~al.}(2006)\citenamefont{Skoge, Donev,
  Stillinger, and Torquato}}]{SkDoStTo06}
\bibinfo{author}{\bibfnamefont{M.}~\bibnamefont{Skoge}},
  \bibinfo{author}{\bibfnamefont{A.}~\bibnamefont{Donev}},
  \bibinfo{author}{\bibfnamefont{F.~H.} \bibnamefont{Stillinger}},
  \bibnamefont{and} \bibinfo{author}{\bibfnamefont{S.}~\bibnamefont{Torquato}},
  \bibinfo{journal}{Phys. Rev. E} \textbf{\bibinfo{volume}{74}},
  \bibinfo{pages}{041127} (\bibinfo{year}{2006}).

\bibitem[{\citenamefont{van Meel et~al.}(2009)\citenamefont{van Meel, Frenkel,
  and Charbonneau}}]{MeFrCh09}
\bibinfo{author}{\bibfnamefont{J.~A.} \bibnamefont{van Meel}},
  \bibinfo{author}{\bibfnamefont{D.}~\bibnamefont{Frenkel}}, \bibnamefont{and}
  \bibinfo{author}{\bibfnamefont{P.}~\bibnamefont{Charbonneau}},
  \bibinfo{journal}{Phys. Rev. E} \textbf{\bibinfo{volume}{79}},
  \bibinfo{pages}{030201 R} (\bibinfo{year}{2009}).

\bibitem[{\citenamefont{Zaccarelli et~al.}(2009)\citenamefont{Zaccarelli,
  Valeriani, Sanz, Poon, Cates, and Pusey}}]{ZaVaSaPoCaPu09}
\bibinfo{author}{\bibfnamefont{E.}~\bibnamefont{Zaccarelli}},
  \bibinfo{author}{\bibfnamefont{C.}~\bibnamefont{Valeriani}},
  \bibinfo{author}{\bibfnamefont{E.}~\bibnamefont{Sanz}},
  \bibinfo{author}{\bibfnamefont{W.~C.~K.} \bibnamefont{Poon}},
  \bibinfo{author}{\bibfnamefont{M.~E.} \bibnamefont{Cates}}, \bibnamefont{and}
  \bibinfo{author}{\bibfnamefont{P.~N.} \bibnamefont{Pusey}},
  \bibinfo{journal}{Phys. Rev. Lett.} \textbf{\bibinfo{volume}{103}},
  \bibinfo{pages}{135704} (\bibinfo{year}{2009}).

\bibitem[{\citenamefont{Charbonneau et~al.}(2010)\citenamefont{Charbonneau,
  Ikeda, van Meel, and Miyazaki}}]{ChIkMeMi10}
\bibinfo{author}{\bibfnamefont{P.}~\bibnamefont{Charbonneau}},
  \bibinfo{author}{\bibfnamefont{A.}~\bibnamefont{Ikeda}},
  \bibinfo{author}{\bibfnamefont{J.~A.} \bibnamefont{van Meel}},
  \bibnamefont{and} \bibinfo{author}{\bibfnamefont{K.}~\bibnamefont{Miyazaki}},
  \bibinfo{journal}{Phys. Rev. E} \textbf{\bibinfo{volume}{81}},
  \bibinfo{pages}{040501(R)} (\bibinfo{year}{2010}).

\bibitem[{\citenamefont{Schmid and Schilling}(2010)}]{ScSc10}
\bibinfo{author}{\bibfnamefont{B.}~\bibnamefont{Schmid}} \bibnamefont{and}
  \bibinfo{author}{\bibfnamefont{R.}~\bibnamefont{Schilling}},
  \bibinfo{journal}{Phys. Rev. E} \textbf{\bibinfo{volume}{81}},
  \bibinfo{pages}{041502} (\bibinfo{year}{2010}).

\bibitem[{\citenamefont{Tarzia}(2007)}]{Tarzia07}
\bibinfo{author}{\bibfnamefont{M.}~\bibnamefont{Tarzia}}, \bibinfo{journal}{J.
  Stat. Mech.} \textbf{\bibinfo{volume}{P01010}} (\bibinfo{year}{2007}).

\bibitem[{\citenamefont{Panagiotopoulos}(2005)}]{Panagiotopoulos05}
\bibinfo{author}{\bibfnamefont{A.~Z.} \bibnamefont{Panagiotopoulos}},
  \bibinfo{journal}{J. Chem. Phys.} \textbf{\bibinfo{volume}{123}},
  \bibinfo{pages}{104504} (\bibinfo{year}{2005}).

\bibitem[{\citenamefont{Fernandes
  et~al.}(2007{\natexlab{a}})\citenamefont{Fernandes, Arenzon, and
  Levin}}]{FeArLe07}
\bibinfo{author}{\bibfnamefont{H.~C.~M.} \bibnamefont{Fernandes}},
  \bibinfo{author}{\bibfnamefont{J.~J.} \bibnamefont{Arenzon}},
  \bibnamefont{and} \bibinfo{author}{\bibfnamefont{Y.}~\bibnamefont{Levin}},
  \bibinfo{journal}{J. Chem. Phys.} \textbf{\bibinfo{volume}{126}},
  \bibinfo{pages}{114508} (\bibinfo{year}{2007}{\natexlab{a}}).

\bibitem[{\citenamefont{Gürsey}(1950)}]{Gursey50}
\bibinfo{author}{\bibfnamefont{F.}~\bibnamefont{Gürsey}},
  \bibinfo{journal}{Proc. Cambridge Philos. Soc.}
  \textbf{\bibinfo{volume}{46}}, \bibinfo{pages}{182} (\bibinfo{year}{1950}).

\bibitem[{\citenamefont{Baxter}(1965)}]{Baxter65}
\bibinfo{author}{\bibfnamefont{R.~J.} \bibnamefont{Baxter}},
  \bibinfo{journal}{Phys. Fluid.} \textbf{\bibinfo{volume}{8}},
  \bibinfo{pages}{687} (\bibinfo{year}{1965}).

\bibitem[{\citenamefont{Lieb and Mattis}(1966)}]{LiMa66}
\bibinfo{author}{\bibfnamefont{E.~H.} \bibnamefont{Lieb}} \bibnamefont{and}
  \bibinfo{author}{\bibfnamefont{D.~C.} \bibnamefont{Mattis}},
  \emph{\bibinfo{title}{Mathematical Physics in one dimension}}
  (\bibinfo{publisher}{Academic Press Inc.}, \bibinfo{address}{New York},
  \bibinfo{year}{1966}).

\bibitem[{\citenamefont{Casey and Runnels}(1969)}]{CaRu69}
\bibinfo{author}{\bibfnamefont{L.}~\bibnamefont{Casey}} \bibnamefont{and}
  \bibinfo{author}{\bibfnamefont{L.}~\bibnamefont{Runnels}},
  \bibinfo{journal}{J. Chem. Phys.} \textbf{\bibinfo{volume}{51}},
  \bibinfo{pages}{5070} (\bibinfo{year}{1969}).

\bibitem[{\citenamefont{Bloomfield}(1970)}]{Bloomfield70}
\bibinfo{author}{\bibfnamefont{V.~A.} \bibnamefont{Bloomfield}},
  \bibinfo{journal}{J. Chem. Phys.} \textbf{\bibinfo{volume}{52}},
  \bibinfo{pages}{2781} (\bibinfo{year}{1970}).

\bibitem[{\citenamefont{Percus}(1976)}]{Percus76}
\bibinfo{author}{\bibfnamefont{J.~K.} \bibnamefont{Percus}},
  \bibinfo{journal}{J. Stat. Phys.} \textbf{\bibinfo{volume}{15}},
  \bibinfo{pages}{505} (\bibinfo{year}{1976}).

\bibitem[{\citenamefont{Percus}(1982)}]{Percus82}
\bibinfo{author}{\bibfnamefont{J.~K.} \bibnamefont{Percus}},
  \bibinfo{journal}{J. Stat. Phys.} \textbf{\bibinfo{volume}{28}},
  \bibinfo{pages}{67} (\bibinfo{year}{1982}).

\bibitem[{\citenamefont{Marko}(1989)}]{Marko89}
\bibinfo{author}{\bibfnamefont{J.}~\bibnamefont{Marko}},
  \bibinfo{journal}{Phys. Rev. Lett.} \textbf{\bibinfo{volume}{62}},
  \bibinfo{pages}{543} (\bibinfo{year}{1989}).

\bibitem[{\citenamefont{Davis}(1990)}]{Davis90}
\bibinfo{author}{\bibfnamefont{H.~T.} \bibnamefont{Davis}},
  \bibinfo{journal}{J. Chem. Phys.} \textbf{\bibinfo{volume}{93}},
  \bibinfo{pages}{4339} (\bibinfo{year}{1990}).

\bibitem[{\citenamefont{Dunlop and Huillet}(2003)}]{DuHu03}
\bibinfo{author}{\bibfnamefont{F.}~\bibnamefont{Dunlop}} \bibnamefont{and}
  \bibinfo{author}{\bibfnamefont{T.}~\bibnamefont{Huillet}},
  \bibinfo{journal}{Physica A} \textbf{\bibinfo{volume}{324}},
  \bibinfo{pages}{698} (\bibinfo{year}{2003}).

\bibitem[{\citenamefont{Fernandes
  et~al.}(2007{\natexlab{b}})\citenamefont{Fernandes, Levin, and
  Arenzon}}]{FeLeAr07}
\bibinfo{author}{\bibfnamefont{H.~C.~M.} \bibnamefont{Fernandes}},
  \bibinfo{author}{\bibfnamefont{Y.}~\bibnamefont{Levin}}, \bibnamefont{and}
  \bibinfo{author}{\bibfnamefont{J.~J.} \bibnamefont{Arenzon}},
  \bibinfo{journal}{Phys. Rev. E} \textbf{\bibinfo{volume}{75}},
  \bibinfo{pages}{052101} (\bibinfo{year}{2007}{\natexlab{b}}).

\bibitem[{\citenamefont{Trizac and Pagonabarraga}(2008)}]{TrPa08}
\bibinfo{author}{\bibfnamefont{E.}~\bibnamefont{Trizac}} \bibnamefont{and}
  \bibinfo{author}{\bibfnamefont{I.}~\bibnamefont{Pagonabarraga}},
  \bibinfo{journal}{Am. J. Phys.} \textbf{\bibinfo{volume}{76}},
  \bibinfo{pages}{777} (\bibinfo{year}{2008}).

\bibitem[{\citenamefont{Freasier and Runnels}(1973)}]{FrRu73}
\bibinfo{author}{\bibfnamefont{B.~C.} \bibnamefont{Freasier}} \bibnamefont{and}
  \bibinfo{author}{\bibfnamefont{L.~K.} \bibnamefont{Runnels}},
  \bibinfo{journal}{J. Chem. Phys.} \textbf{\bibinfo{volume}{58}},
  \bibinfo{pages}{2963} (\bibinfo{year}{1973}).

\bibitem[{\citenamefont{Fulińiski and Longa}(1979)}]{FuLo79}
\bibinfo{author}{\bibfnamefont{A.}~\bibnamefont{Fulińiski}} \bibnamefont{and}
  \bibinfo{author}{\bibfnamefont{L.}~\bibnamefont{Longa}}, \bibinfo{journal}{J.
  Stat. Phys.} \textbf{\bibinfo{volume}{21}}, \bibinfo{pages}{635}
  (\bibinfo{year}{1979}).

\bibitem[{\citenamefont{Ben-Naim and Krapivsky}(2006)}]{BeKr06}
\bibinfo{author}{\bibfnamefont{E.}~\bibnamefont{Ben-Naim}} \bibnamefont{and}
  \bibinfo{author}{\bibfnamefont{P.~L.} \bibnamefont{Krapivsky}},
  \bibinfo{journal}{Phys. Rev. E} \textbf{\bibinfo{volume}{73}},
  \bibinfo{pages}{031109} (\bibinfo{year}{2006}).

\bibitem[{\citenamefont{Kantor and Kardar}(2009{\natexlab{a}})}]{KaKa09a}
\bibinfo{author}{\bibfnamefont{Y.}~\bibnamefont{Kantor}} \bibnamefont{and}
  \bibinfo{author}{\bibfnamefont{M.}~\bibnamefont{Kardar}},
  \bibinfo{journal}{Phys. Rev. E} \textbf{\bibinfo{volume}{79}},
  \bibinfo{pages}{041109} (\bibinfo{year}{2009}{\natexlab{a}}).

\bibitem[{\citenamefont{Kantor and Kardar}(2009{\natexlab{b}})}]{KaKa09b}
\bibinfo{author}{\bibfnamefont{Y.}~\bibnamefont{Kantor}} \bibnamefont{and}
  \bibinfo{author}{\bibfnamefont{M.}~\bibnamefont{Kardar}},
  \bibinfo{journal}{EPL (Europhysics Letters)} \textbf{\bibinfo{volume}{87}},
  \bibinfo{pages}{60002} (\bibinfo{year}{2009}{\natexlab{b}}).

\bibitem[{\citenamefont{Gurin and Varga}(2010)}]{GuVa10}
\bibinfo{author}{\bibfnamefont{P.}~\bibnamefont{Gurin}} \bibnamefont{and}
  \bibinfo{author}{\bibfnamefont{S.}~\bibnamefont{Varga}},
  \bibinfo{journal}{Phys. Rev. E} \textbf{\bibinfo{volume}{82}},
  \bibinfo{pages}{041713} (\bibinfo{year}{2010}).

\bibitem[{\citenamefont{Bowles}(2000)}]{Bowles00}
\bibinfo{author}{\bibfnamefont{R.~K.} \bibnamefont{Bowles}},
  \bibinfo{journal}{Physica A} \textbf{\bibinfo{volume}{275}},
  \bibinfo{pages}{217} (\bibinfo{year}{2000}).

\bibitem[{\citenamefont{Stadler et~al.}(2002)\citenamefont{Stadler, Luck, and
  Mehta}}]{StLuMe02}
\bibinfo{author}{\bibfnamefont{P.~F.} \bibnamefont{Stadler}},
  \bibinfo{author}{\bibfnamefont{J.~M.} \bibnamefont{Luck}}, \bibnamefont{and}
  \bibinfo{author}{\bibfnamefont{A.}~\bibnamefont{Mehta}},
  \bibinfo{journal}{Europhys. Lett.} \textbf{\bibinfo{volume}{57}},
  \bibinfo{pages}{46} (\bibinfo{year}{2002}).

\bibitem[{\citenamefont{Stinchcombe and Depken}(2002)}]{StDe02}
\bibinfo{author}{\bibfnamefont{R.}~\bibnamefont{Stinchcombe}} \bibnamefont{and}
  \bibinfo{author}{\bibfnamefont{M.}~\bibnamefont{Depken}},
  \bibinfo{journal}{Phys. Rev. Lett.} \textbf{\bibinfo{volume}{88}},
  \bibinfo{pages}{125701} (\bibinfo{year}{2002}).

\bibitem[{\citenamefont{Bowles and Saika-Voivod}(2006)}]{BoSa06}
\bibinfo{author}{\bibfnamefont{R.~K.} \bibnamefont{Bowles}} \bibnamefont{and}
  \bibinfo{author}{\bibfnamefont{I.}~\bibnamefont{Saika-Voivod}},
  \bibinfo{journal}{Phys. Rev. E} \textbf{\bibinfo{volume}{73}},
  \bibinfo{pages}{011503} (\bibinfo{year}{2006}).

\bibitem[{\citenamefont{Dhar and Lebowitz}(2010)}]{DhLe10}
\bibinfo{author}{\bibfnamefont{D.}~\bibnamefont{Dhar}} \bibnamefont{and}
  \bibinfo{author}{\bibfnamefont{J.~L.} \bibnamefont{Lebowitz}},
  \bibinfo{journal}{EPL} \textbf{\bibinfo{volume}{92}}, \bibinfo{pages}{20008}
  (\bibinfo{year}{2010}).

\bibitem[{\citenamefont{Fredrickson and Andersen}(1984)}]{FrAn84}
\bibinfo{author}{\bibfnamefont{G.~H.} \bibnamefont{Fredrickson}}
  \bibnamefont{and} \bibinfo{author}{\bibfnamefont{H.~C.}
  \bibnamefont{Andersen}}, \bibinfo{journal}{Phys. Rev. Lett.}
  \textbf{\bibinfo{volume}{53}}, \bibinfo{pages}{1244} (\bibinfo{year}{1984}).

\bibitem[{\citenamefont{Kob and Andersen}(1993)}]{KoAn93}
\bibinfo{author}{\bibfnamefont{K.}~\bibnamefont{Kob}} \bibnamefont{and}
  \bibinfo{author}{\bibfnamefont{H.~C.} \bibnamefont{Andersen}},
  \bibinfo{journal}{Phys. Rev. E} \textbf{\bibinfo{volume}{48}},
  \bibinfo{pages}{4364} (\bibinfo{year}{1993}).

\bibitem[{\citenamefont{Ritort and Sollich}(2003)}]{RiSo03}
\bibinfo{author}{\bibfnamefont{F.}~\bibnamefont{Ritort}} \bibnamefont{and}
  \bibinfo{author}{\bibfnamefont{P.}~\bibnamefont{Sollich}},
  \bibinfo{journal}{Adv. Phys.} \textbf{\bibinfo{volume}{52}},
  \bibinfo{pages}{219} (\bibinfo{year}{2003}).

\bibitem[{\citenamefont{Dhar}(2002)}]{Dhar02}
\bibinfo{author}{\bibfnamefont{D.}~\bibnamefont{Dhar}}, \bibinfo{journal}{Phys.
  A} \textbf{\bibinfo{volume}{5}}, \bibinfo{pages}{315} (\bibinfo{year}{2002}).

\bibitem[{\citenamefont{Onsager}(1949)}]{Onsager49}
\bibinfo{author}{\bibfnamefont{L.}~\bibnamefont{Onsager}},
  \bibinfo{journal}{Ann. N. Y. Acad. Sci.} \textbf{\bibinfo{volume}{51}},
  \bibinfo{pages}{627} (\bibinfo{year}{1949}).

\bibitem[{\citenamefont{Villarruel et~al.}(2000)\citenamefont{Villarruel,
  Lauderdale, Mueth, and Jaeger}}]{ViLaMuJa00}
\bibinfo{author}{\bibfnamefont{F.~X.} \bibnamefont{Villarruel}},
  \bibinfo{author}{\bibfnamefont{B.~E.} \bibnamefont{Lauderdale}},
  \bibinfo{author}{\bibfnamefont{D.~M.} \bibnamefont{Mueth}}, \bibnamefont{and}
  \bibinfo{author}{\bibfnamefont{H.~M.} \bibnamefont{Jaeger}},
  \bibinfo{journal}{Phys. Rev. E} \textbf{\bibinfo{volume}{61}},
  \bibinfo{pages}{6914} (\bibinfo{year}{2000}).

\end{thebibliography}

\end{document}